# Real-space Visualization of Charge Density Wave Induced Local Inversion-Symmetry Breaking in a Skyrmion Magnet

Haoyang Ni[1,2], William R. Meier[3], Hu Miao[4], Andrew J. May[4], Brian C. Sales[4], Jian-min Zuo[1,#], Miaofang Chi[2,#]

[1]Department of Materials Science and Engineering, University of Illinois at Urbana-Champaign, Urbana, Illinois 61801, USA
[2]Center for Nanophase Materials Sciences, Oak Ridge National Laboratory, Oak Ridge, TN 37831, USA
[3]Department of Materials Science and Engineering, University of Tennessee, Knoxville, 37996, USA
[4]Materials Science and Technology Division, Oak Ridge National Laboratory, Oak Ridge, Tennessee, 37831, USA

#Correspondence should be addressed to chim@ornl.gov; jianzuo@illinois.edu

**Intertwining charge density wave (CDW) with spin and pairing order parameters is a major focus of contemporary condensed matter physics[1]. Lattice distortions and local symmetry breaking resulted from CDWs are crucial for the emergence of correlated and topological states in quantum materials in general[2]. While the presence of CDWs can be detected by diffraction or spectroscopic techniques, atomic visualization of the CDW induced lattice distortions remains limited to CDW with short wavelengths. In this letter, we realized the imaging of incommensurate long-wavelength CDWs based on cryogenic four-dimensional scanning transmission electron microscopy (cryo-4DSTEM). By visualizing the incommensurate CDW induced lattice modulations in a skyrmion magnet $EuAl_4$, we discover two out-of-phase intra-unit cell shear modulations that specifically break the local inversion-symmetry. Our results provide direct evidence for the intertwined spin and charge orders in $EuAl_4$ and key information about local symmetry. Furthermore, we establish cryo-4DSTEM as an indispensable approach to understand CDW induced new quantum states of matter.**

CDW, an electronic fluid that breaks the translational symmetry, has been one of the central focuses in condensed matter physics over seven-decades[3,4]. Microscopically, CDW changes both electronic and lattice structures and hence exhibit macroscopic physical consequences. Recently, CDWs are found to play a pivotal role in the intertwined and topological materials, where CDWs couple with superconductivity, pair density wave, three-dimensional quantum Hall effect (3D-QHE) and axionic electrodynamics, etc.[2,5–7] Fundamental understanding of these new quantum states of matters requires experimental identification of CDW induced lattice distortions and local symmetry breakings.

Lattice structures in the CDW phase are traditionally determined through diffraction techniques using X-rays or neutrons[8,9], which provide useful information on the average crystal structure, but could overlook local details due to limited spatial resolution. Alternatively, scanning tunneling microscopy (STM) provides high spatial resolutions to imaging CDWs, but information is limited on the surface that may or may not represent the intrinsic bulk properties[10]. More recently, high-resolution scanning transmission electron microscopy (STEM) imaging at cryogenic temperatures has become available with the improved stability of low temperature holders, which allows direct visualization of atomic displacements in CDW materials at pm precision, but its applications are still limited to the CDW phases with small wavelengths[11–14]. Real space imaging of CDW materials with more than 5 to 10 nm wavelengths with larger length scales statistics and dynamics at cryogenic temperature is still challenging. it is especially challenging to analyze local lattice distortions in incommensurate, long wavelength CDWs where local lattice distortion could be very small, in the order of pm.

Among unidirectional, long-wavelength CDW materials, EuAl$_4$ features a simple tetragonal crystal structure at room temperature[15] (Fig. 1a). Below 140 K, EuAl$_4$ hosts a prototypical incommensurate CDW (ICDW) with a small CDW wavevector along the crystal c-axis and (Fig. 1f), corresponding to a long wavelength of $\lambda_{CDW} \sim 6.5\ nm$ [15–17]. However, the microscopic origin of this long-wavelength CDW and its interplay with magnetic skyrmions in EuAl$_4$ below Neél temperature (15.4 K) are yet to be understood[18–20]. Previous research, based on a centrosymmetric CDW structure, linked the noncollinear spin states in EuAl$_4$ to an electronically driven Ruderman-Kittel-Kasuya-Yosida (RKKY) interaction, citing the structural similarities to GdRu$_2$Si$_2$[21,22]. However, quantitatively explaining multi-Q spin states with RKKY remains challenging[18,19,23]. There are also doubts due to the differing spin and charge modulation directions between the two materials[19,24]. These uncertainties highlight the need to revisit the lattice structure and symmetry in the CDW of EuAl$_4$ to fully understand its complex magnetic phase diagram.

In this study, by analyzing the convergent beam electron diffraction (CBED) patterns captured in the cryo-4DSTEM datasets, we can quantitatively identify the local inversion symmetry breaking and to map atomic displacements in the real space, which represents the first realization of visualizing a long wavelength CDW in the bulk state. Using this method, we observed the CDW-induced local inversion symmetry breaking in the presence of global inversion symmetry. The CDW could be described by a combination of two intra-unit cell modulations of Al-Al distances at the Al1 and Al2 sites (Fig. 1a) with a $0.64\pi$ phase difference between them, yielding an unconventional lattice deformation. This structural insight provides valuable and detailed local information for the theoretical studies for EuAl$_4$. Furthermore, we demonstrate that cryo-4DSTEM as a powerful new probe for CDW with potential for the study of other intertwined orders in correlated quantum materials.

## Results

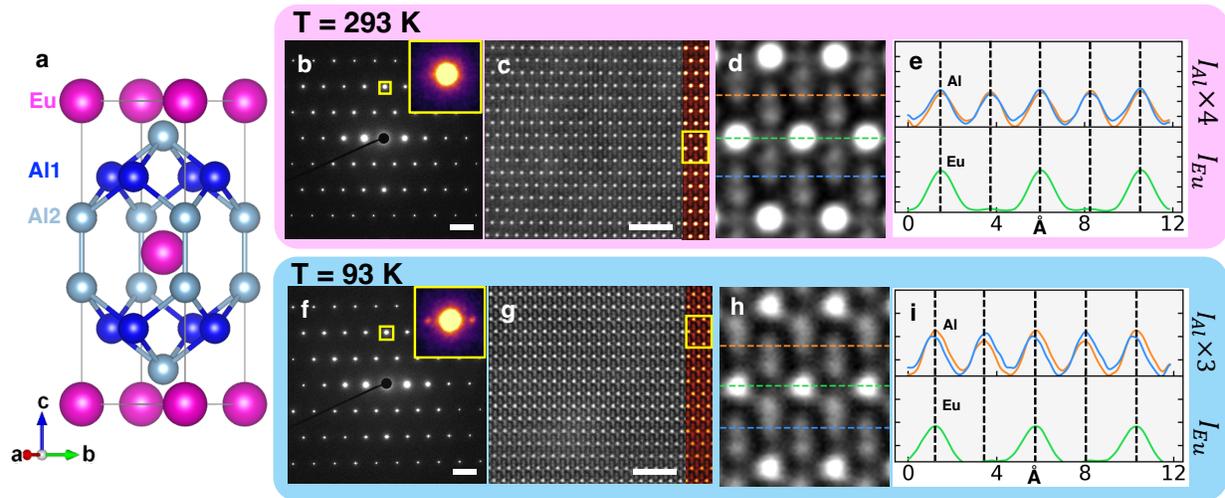

**Fig. 1, Observation of CDW induced lattice modulations in EuAl$_4$.** *a, Room-temperature structure of EuAl4 b, f, typical SAED pattern of EuAl$_4$ along the [100] axis. Insets are enlarged (020) reflection. Inset in e shows satellite spots with 0.172 r.l.u. Scale bar 5 mrad. c, g, STEM images of EuAl$_4$ at 93K and 293K, along with horizontally averaged unit cell image, recorded*

*using an annular-bright-field detector. Scale bar 2 nm. Contrast inverted for visualization **d, h**, enlarged unit cell image from **c, g,** and **e, i,** line profiles showing Al1-Al1 interatomic distances (orange and blue), compared with Eu-Eu interatomic distances (green). Supercell images were generated by averaging along b direction and shown beside full-frame image to further enhance the structural details.*

We start by establishing the presence of CDW in EuAl$_4$ using selected area electron diffraction (SAED). Fig. 1b,f show the SAED patterns of EuAl$_4$ along $[100]$ recorded at the temperatures of 293 K and 93 K, respectively. At 93 K, distinct satellite spots at $q = 0.172c^*$ were observed corresponding to $\lambda_{CDW} \sim 6.5\ nm$. Atomic-resolution STEM images of EuAl$_4$ at these temperatures are shown in Fig. 1b-e and g-i, revealing both Eu and Al atomic columns projected along $[100]$. The line profiles of the inverted-contrast images show two Al1 chains and the Eu chain (Fig. 1d and 1h). By utilizing Eu atoms as the reference points, the Al atomic displacements in the CDW phase can be detected (Fig. 1h), indicating dimerization between nearest-neighboring Al1 atoms. The contrast of Al atoms is weak and the field of view of atomic-resolution imaging is restricted, which makes it difficult to determine the CDW symmetry and structure with statistical certainty.

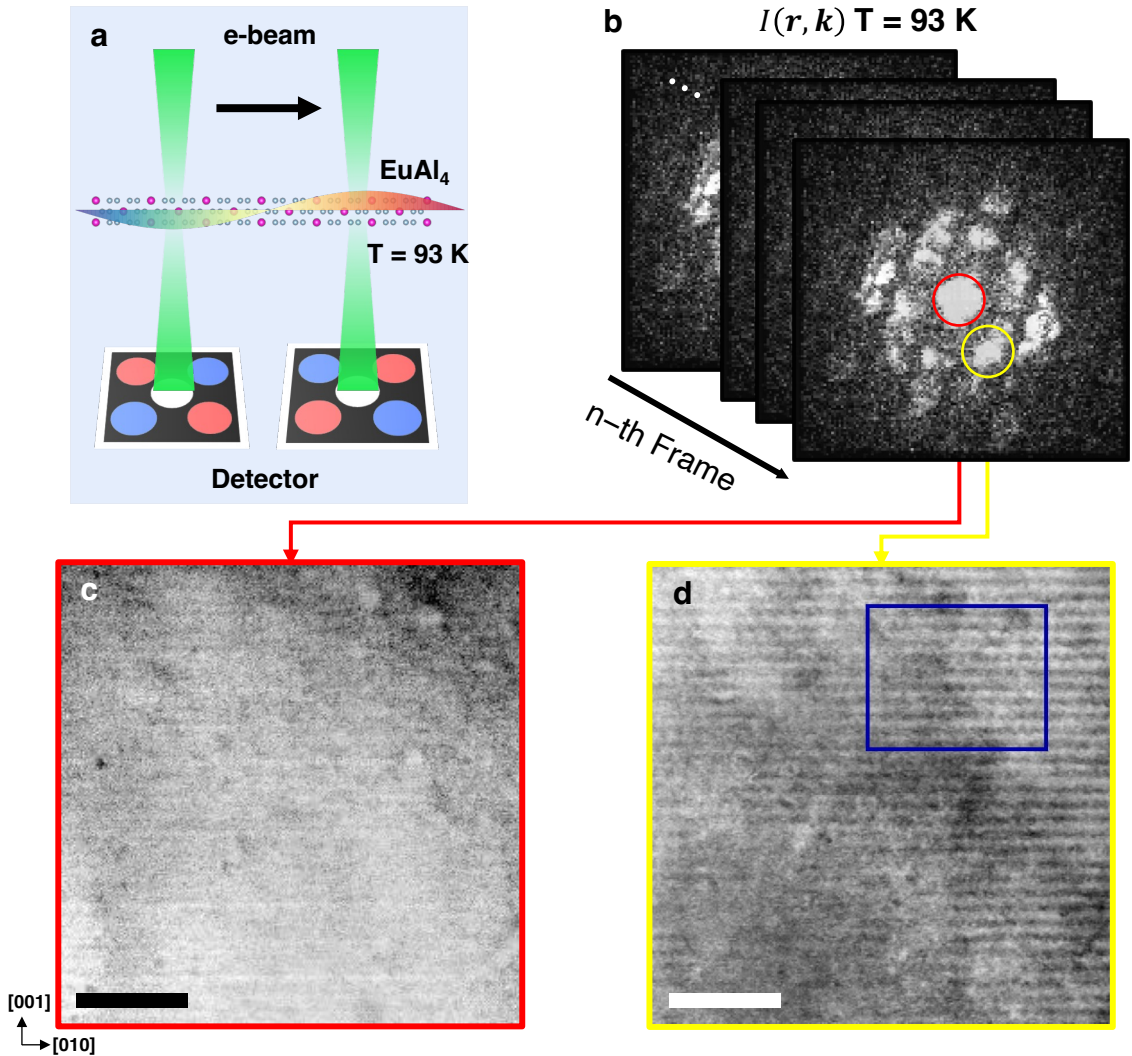

**Fig. 2, cryo-4D-STEM. *a*,** *Schematics of SCBED. Electron beam raster samples and collect CBED pattern at each probe position.* ***b***, *Representative SCBED dataset acquired from experimental setup shown in **a**. Direct transmitting beam and reflection $(01\bar{3})$ is circled for reconstructing virtual bright field (VBF) (c) and dark field images (VDF) (d) images at 93 K. At 93 K, VDF image shows well-defined fringes while VBF image does not. ROI labeled by blue box in **d** is most on zone axis and CBED patterns in this region are analyzed and discussed further. Scale bar 50 nm for **c, d**.*

To improve the electron beam sensitivity to CDWs, we resort to cryo-4DSTEM in CBED mode, which is schematically depicted in Fig. 2a. This technique collects CBED patterns by scanning the electron probe across a region of interest, resulting in a 4D datasets, with the dimensional coordinates of the probe position $(x, y)$ and the diffraction coordinate $(k_x, k_y)$[25]. CBED is an aberration-free technique in reciprocal space that offers sensitivity to picometer-scale atomic displacement[26–28]. CBED combined with 4DSTEM enables the determination of local crystal symmetry[28–30], lattice distortions and strain[29,30].

Figures 2**c** and **d** show the reconstructed bright field and dark field images from the 4DSTEM dataset, respectively, at a temperature of 93 K. The 4DSTEM datasets here were acquired with a field-of-view of 202.4 nm and 0.8 nm sampling with a spatial resolution of ~1 nm. In Fig. 2, the dark field image is reconstructed using the intensity of $(01\bar{3})$, which reveals distinct fringe features with a periodicity of approximately 6.5 nm (further details in Supplementary Fig. 2). The $(01\bar{3})$ reflection was chosen due to its high sensitivity to the CDW structural modulations (see Supplementary Note 2). In comparison, the reconstructed bright-field image (Fig. 2c) displays relatively uniform contrast. This is because central beam intensity is not sensitive to atomic displacements (Supplementary Note 2). Furthermore, VDF images reconstructed at T = 293 K (Supplementary Fig. 5a) does not show any fringes. Combining these observations, we can confirm that the fringe pattern observed in VDF images originate from the CDW. Using 4DSTEM, we are allowed to probe CDW with large field-of-view while maintaining sensitivity to minute atomic displacement, which addresses the challenges we encountered in atomic-resolution STEM imaging.

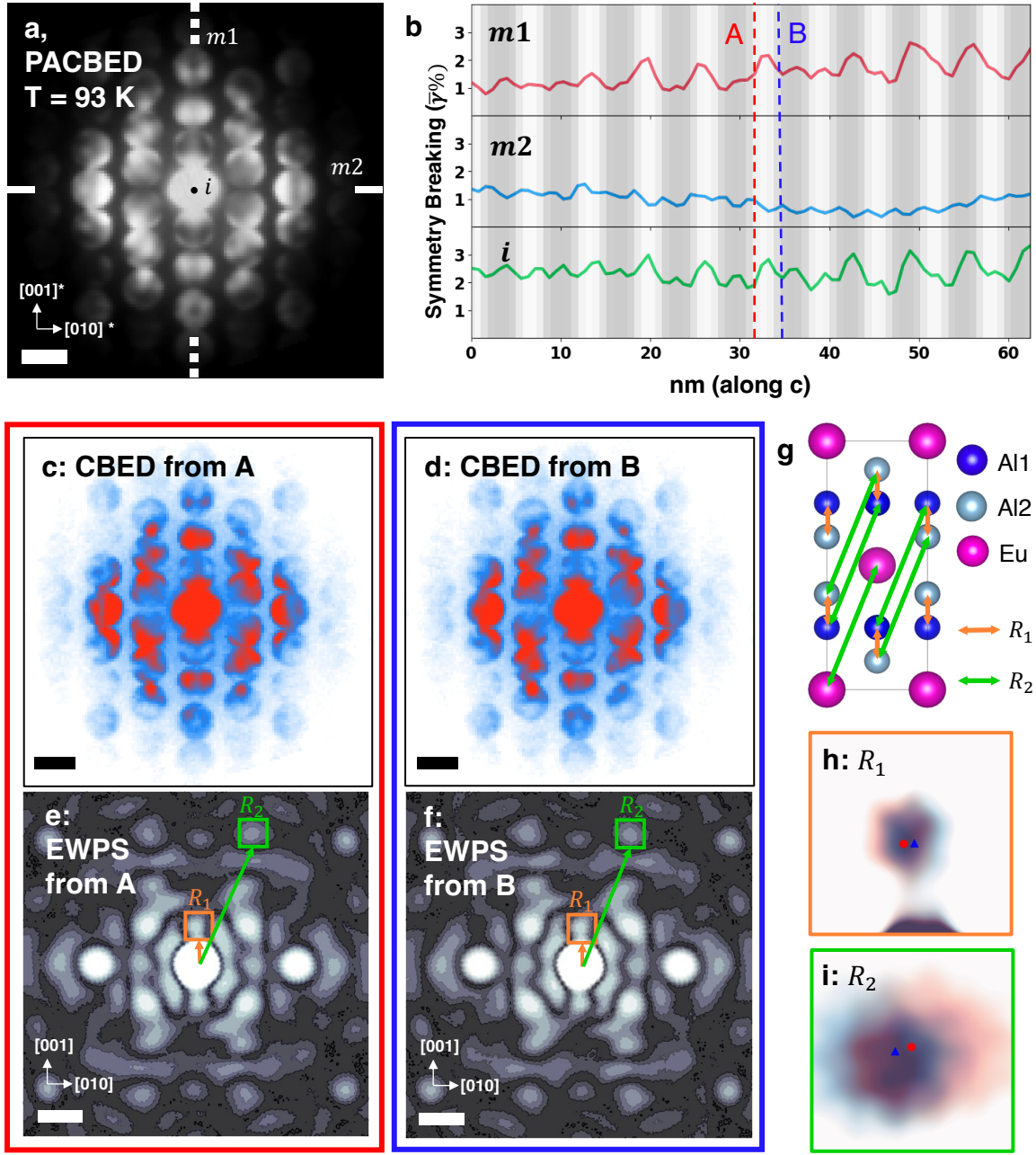

**Fig. 3a**, Position-averaged convergent beam electron diffraction (PACBED) from the region labeled in Fig. 2d, with symmetry elements to be quantified m1, m2, i labeled. Scale bar 5 mrad. **b**, Quantification of symmetry breaking of CBED pattern along **c** axis, by defining symmetry breaking factor $\bar{\gamma}\% = 100\% - \gamma\%$, where $\gamma$ represents the normalized cross-correlation value between symmetry-related intensity feature. $\bar{\gamma}\%$ curves are overlaid on VDF fringe intensity, showing the fluctuation has a periodicity half of the CDW wavelength. **c, d**, CBED patterns from position A and B, and **e, f** their corresponding EWPS. Two peaks labeled in orange and green correspond to interatomic

*vectors $R_1$ and $R_2$. Atomic pairs related to $R_1$ and $R_2$ are labeled in EuAl₄ model in **g**. The two peaks are overlaid on each other to visualize the peak shift in **h, i**. Scale bar 5 mrad for **c, d**, 2Å for **e, f**.*

We then quantify the symmetry of each CBED pattern along the ***c***-axis in the 4DSTEM dataset[31,32] (Supplementary Note 3) acquired at T = 93 K. Above the CDW transition temperature ($T_{CDW}$), EuAl₄ has a space group $I4/mmm$ symmetry. Under this symmetry, zone-axis CBED pattern along [100], exhibits the $2mm$ symmetry with two perpendicular mirror planes along [001]*, [010]*, and an inversion center[33] (labelled as $m1, m2, i$ in Fig. 3a and Supplementary Fig. 5). Below the $T_{CDW}$, the CBED pattern summed over ~10 CDW periods reveals the highest diffraction pattern symmetry to be $m$. According to Buxton et al., the possible crystal point groups include tetragonal $4/m$, $4mm$ or orthorhombic $mmm$[33]. Since the preserved mirror plane is perpendicular to [001], both $4/m$ and $mmm$ are feasible. Both possible point groups are centrosymmetric, proving the global centrosymmetric structure below $T_{CDW}$.

For individual CBED patterns, all three pattern symmetries, i.e. $m1, m2, and\ i$, of the CDW phase, however, show symmetry-breaking that is quantified by the breaking factor $\bar{\gamma} = 100 - \gamma$. Here $\gamma$ is the normalized cross-correlation between symmetry-related intensity features in CBED patterns in percentage. Therefore, a larger $\bar{\gamma}$ indicates a greater deviation from perfect symmetry. As shown in Fig. 3b, $\bar{\gamma}$ is determined as 1% to 3% for $m1$ and $i$, respectively, and approximately 0.5% to 1% for $m2$. At the highest symmetry breaking point, the CBED pattern symmetry reduces to 1. The highest possible symmetry include tetragonal 4, $\bar{4}$ and orthorhombic 222, $mm2$[33], all of which are *non*-centrosymmetric. For all three symmetry elements $\bar{\gamma}$ fluctuates with a periodicity half of the CDW wavelength. This periodicity is evident from the FFT of the $\bar{\gamma}$ curves in Fig. 3**b,** which yields a value of 0.31 nm⁻¹ (Supplementary Fig. 6), corresponding to a real-space periodicity of 3.23 nm, e.g., half of the CDW periodicity. Interestingly, the mirror symmetry $m2$ breaks in an opposite way compared to the inversion symmetry $i$ (Fig. 3b). This symmetry breaking is directly observable in the CBED patterns. Fig. 3**c and d** depict the CBED patterns obtained from two positions (A and B in Fig. 3**b**) separated by half of the CDW period, i.e. $\pi$ phase distance. Although CBED patterns at position A and B are both quantified to be more centrosymmetric, their intensity features are mirrored. This result thus reveals not only the presence of local symmetry breaking but also a phase shift between the breaking of inversion and mirror symmetry within the EuAl4 CDW.

We calculated the exit wave power spectra (EWPS) of CBED patterns to investigate the origin of symmetry breaking within the CDW phase (Supplementary Note 4). Fig. 3e and 3f are two examples. The peak positions in EWPS carry information of interatomic vectors, which can be well used to determine lattice distortion variations in real space (Supplementary Note 5). The EWPS peaks marked with $R_1$ in Figure 3e and 3f, for example, correspond to the interatomic vector of the Al1-Al2 interatomic vector (indicated as vector $R_1$ in Fig. 3g). The ones marked with $R_2$ corresponds to the Al-Al/Eu-Eu interatomic vector equal to half of sum of the unit cell lattice [010] and [001]. (vector $R_2$ in Fig. 3g).

The EWPS peaks of $R_1$ and $R_2$ show a horizontal periodical movement as the beam scans across c-direction (Supplementary Video 2). The movements indicate a displacement perpendicular to the CDW modulation direction [001], supporting a transverse CDW, consistent with the x-ray scattering measurement[34]. This transverse modulation is also observed by overlaying the EWPS peak positions of $R_1$ and $R_2$, respectively, as shown in Fig. 3h and 3i, as the EWPS peaks are separated along [010]. Interestingly, these two vectors displace in opposite directions, i.e., $R_1$ displaces towards [0$\bar{1}$0] direction while $R_2$ moves towards the [010] direction. This behavior suggests an out-of-phase modulations between these two interatomic vectors.

Quantitative measurements of the peak positions revealed unexpected structural modulations. Figures 4**a** and **b** map the horizontal peak shift, $\Delta x$, of $R_1$ and $R_2$ peaks. For both peaks, the $\Delta x$ maps show a wave-like feature. By fitting the averaged $\Delta x$ along **b** to a sine function $\Delta x = A_x \sin\left(\frac{2\pi x}{\lambda} + \varphi\right)$, the modulation amplitudes were determined to be 2.47 pm for $R_1$ and 3.52 pm for $R_2$, respectively. The $\lambda$ in both cases is ~6.5 nm, which is same as the observed CDW periodicity by SAED. The phase, $\varphi$, is surprisingly different with the $R_1$ peak modulation exhibits a leading phase of $0.64\pi$ relative to $R_2$. Along **c,** the measured $\Delta y$ shows no periodicity in the peak displacement maps (Supplementary Fig. 9). These results further confirm the transverse CDW that the atomic displacement is dominantly perpendicular to [001].

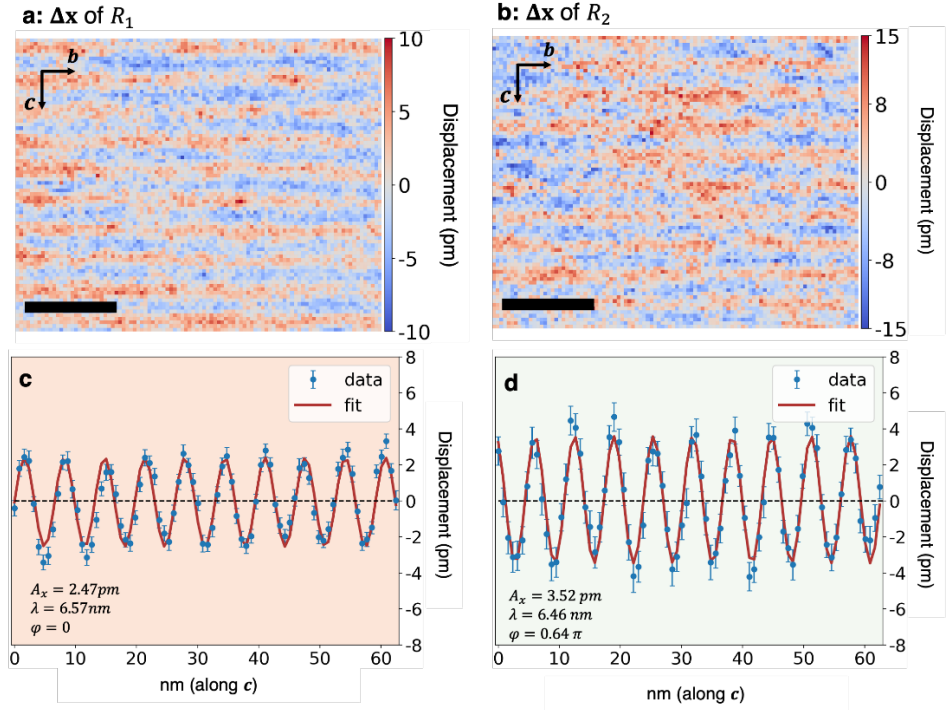

**Fig. 4 *a, b,*** *Quantitative 2D mapping of displacement along **a** for EWPS peaks $R_1$ and $R_2$. **c, d,** are the experimental and fitting of peak modulation from **a, b,** respectively. The modulation shows a good fit to a sinusoidal function for both $R_1$ and $R_2$ displacements. The peak positions here were determined using the center-of-mass method. Scale bar is 20 nm. Error bar shows a 95% confidence interval along **a**.*

**Atomic Displacement Model**

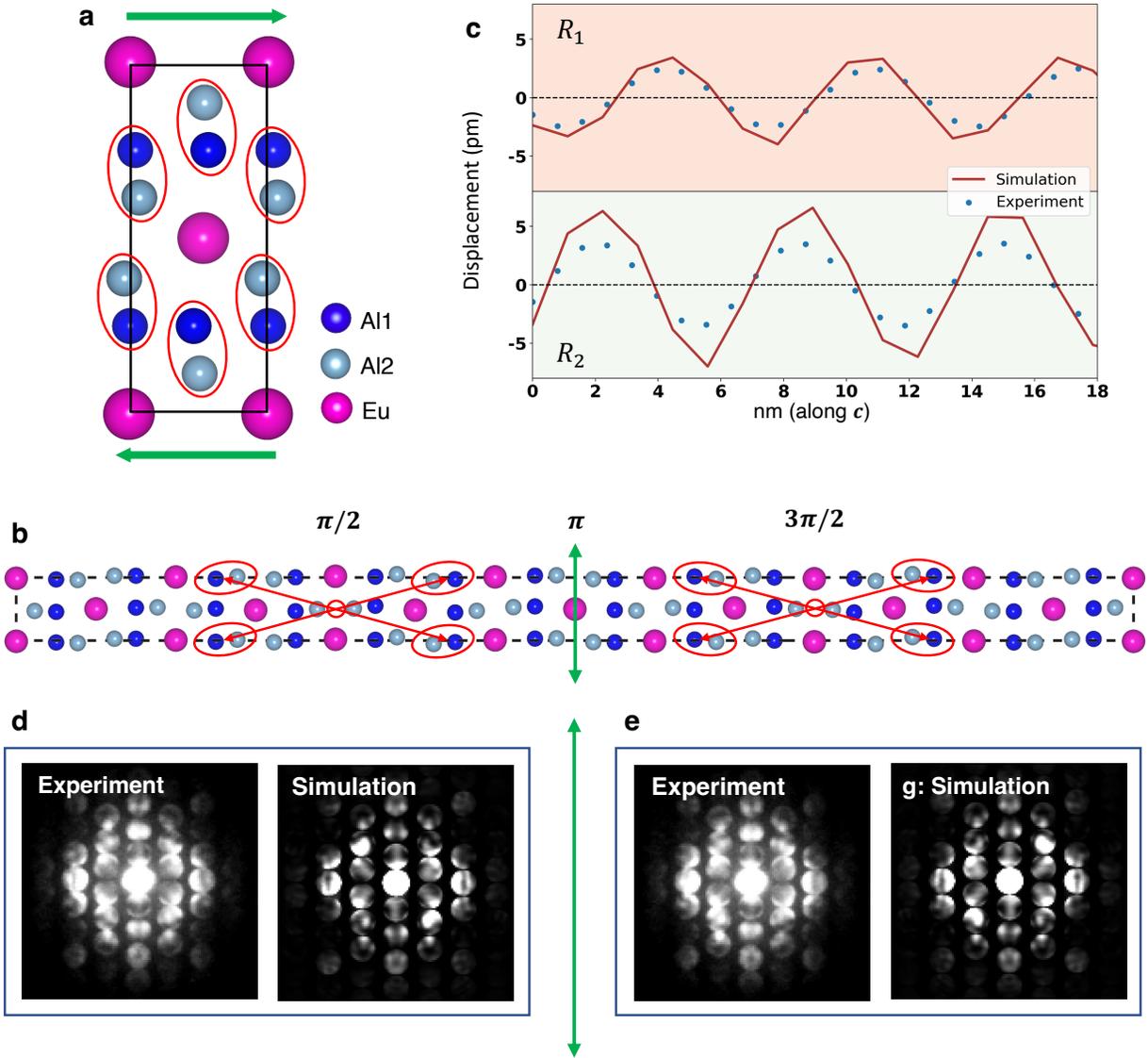

**Fig. 5** *a, The atomic displacement model in EuAl$_4$. **a** Dimerization and shearing of Al atoms. The green arrows indicate yz-shearing and red ellipses indicate pairs of nearest neighbor Al atoms and their relative shift. **b**, Atomic structure of the proposed EuAl$_4$ CDW over 1 CDW period. Approximate mirror plane (green arrow) and inversion centers (open red circles) are labeled. **c**, the modulation of R$_1$ and R$_2$ peak positions from simulated EWPS (red solid line) plotted against the experimental results (blue dots). **d, e**, pairs of experimental and simulated CBED patterns with π phase away, related by mirror plane. Their match validates the proposed structure model. Displacement of atoms in a,b are exaggerated by a factor of 2 for visualization.*

Fig. 5a qualitatively presents the atomic model of one CDW period derived from the experimentally observed $R_1$ and $R_2$ modulations, with the center Eu atom used as the reference. This model includes the modulation of $R_2$ which is mainly contributed by the $yz$-shearing of unit-cell (green arrows), while $R_1$ arising from the distortion between Al1-Al2 pairs (red ellipses), where Al1 and Al2 move in an opposite direction relative to each other. Such modulation of Al1-Al2 distance explains the dimerization observed in STEM images (Fig. 1**e-h**). Integrating the quantified phase and amplitude of the two modulation functions, the CDW structure is deduced as depicted in Fig. 5b.

To validate our proposal, we simulated CBED and EWPS based on the proposed model (Supplementary Note 7) and calculated the corresponding modulations of $R_1$ and $R_2$. Fig. 5c compares the simulated and experimentally obtained modulations of $R_1$ and $R_2$, demonstrating a good consistency between them in both phase and amplitude. Fig. 5d, e shows two pairs of simulated and experimental CBED patterns with the probe position shifted by $\pi$ in phase. Both the simulated and experimental patterns exhibit the 'mirroring' phenomenon observed in the experiments (Fig. 3**c, d**), and the periodically modulating CBED and EWPS patterns from Supplementary Video 3, 4. From Fig. 5b, we see that when center Eu atom coincides with CDW phase $\pi$, approximate mirror planes appear at $\pi$ (green) and inversion centers (red) at $\pi/2$ or $3\pi/2$, which is consistent with our observation of symmetry fluctuation shown in Fig. 3b. These results explicate the multiple symmetry-breaking features observed experimentally.

The unconventional atomic displacements involving two different modulations with different phase fronts revealed for the EuAl$_4$ supercell structure have been overlooked in previous studies. The superspace group $Fmmm(00g)s00$ derived from X-ray results[34] suggests the preservation of overall crystal inversion symmetry and transverse atomic displacements in EuAl$_4$ CDW phase, which was also observed here using CBED. However, the local displacement modulations suggested by X-ray diffraction differ from our model. In the X-ray model, the modulation of $R_1$ and $R_2$ peaks are in-phase (Supplementary Fig. 12), rather than out-of-phase as shown in our observation. Further, the symmetry-equivalent pairs of Al atoms in our structure exhibit different modulation amplitudes (Supplementary Fig. 9, 11), in contrast to the X-ray model where they are constrained to be the same. These difference could stem from assumptions that were made in the interpretation of X-ray experimental patterns, including the assumption of certain group-subgroup relations and orthorhombic twinning[34]. While our model assumes only structural homogeneity along the incident beam direction.

Our observation of multiple structural modulation modes in the EuAl$_4$ CDW phase offers critical insights into the role of the CDW in stabilizing multi-Q spin states. We reveal that the out-of-plane CDW disrupts local inversion symmetry due to out-of-phase lattice modulations, invalidating the previously assumed centrosymmetry in theoretical studies. This local inversion symmetry-breaking and large CDW unit cell allows Dzyaloshiinski-Moriya interaction (DMI)[35] in the ab-plane and weakly coupled along the c-axis.

To summarize, we have unraveled unexpected atomic displacement modulations within the complex long-wavelength I-CDW phase of EuAl$_4$. The two modulation modes; both exhibit sinusoidal wave-like features with different phase fronts, leading to an unconventional local inversion symmetry breaking from lattice deformations. The structure model established in this work provides new refinements for future theoretical calculations, potentially providing new insights for microscopic mechanism of the formation of low temperature spin states. The unraveling of the CDW structure was enabled by direct visualization of local symmetry and measurement of local atomic displacements within the complex long-wavelength CDW at pm precision, using cryo-4DSTEM and EWPS analysis. The methodology advanced here can serve as an indispensable method for the general study of CDW and intertwined orders for correlated quantum materials.

## Methods:

### Crystal growth and Sample Preparation

EuAl$_4$ crystal was grown using Al self-flux method, the same as in previous works[20]. TEM specimen was prepared by mechanical polishing and ion milling for thinning.

### Atomic Resolution STEM Imaging

Electron microscopy and diffraction experiments were done on JEOL JEM-ARM200F (NeoARM) aberration-corrected scanning transmission electron microscope (STEM) operating at

200 kV. Cryo experiments were carried out using Gatan 636 double tilt cooling holder, with base temperature at 92.5 K. Atomic resolution STEM images were acquired with 27.4 mrad convergence angle with 50 pA beam current measured by Faraday cup. Collection angle is 12-24 mrad for annular bright field (ABF). At 93 K, low signal-to-noise ratio (SNR) image (Supplementary. Fig. 1) series containing 30 frames with 1 $\mu s$/pixel dwell time and 0.19Å/pixel sampling. Frame size is $512 \times 512$ pixels. Image distortion due to thermal instability and bubbling is still observable in single frame images despite short acquisition time. For this reason, image series were non-rigid registered using PyMatchSeries package[36,37] to produce high SNR STEM images. The resultant ABF image (Supplementary Fig. 1) shows a maximum information transfer of 0.95Å.

**Scanning CBED**

Scanning convergent beam electron diffraction (SCBED) was acquired at 200 kV with convergence angle of ~2.2 mrad. 4D-STEM datasets were acquired using PNDetector[38], operating at 1,000 fps readout rate with FoV of 202.4 nm with 0.79 nm/pixel sampling.

**Acknowledgements**

The microscopy work was supported by an Early Career project supported by DOE Office of Science FWP #ERKCZ55–KC040304. All microscopy technique development was performed and supported by Oak Ridge National Laboratory's (ORNL) Center for Nanophase Materials Sciences (CNMS), which is a DOE Office of Science User Facility. JMZ is supported by the U.S. DOE, Office of Science, Basic Energy Sciences, Materials Sciences and Engineering Division under Contract No. DE-SC0022060.